# Prompt identification of solar wind stream interaction regions from Survey Burst Mode observations of the Radio and Plasma Wave experiment on Solar Orbiter

Dmytro Chechotkin[1*], Oleksiy Dudnik[2,1], Oleksandr Yakovlev[1]

[1]Institute of Radio Astronomy of the National Academy of Sciences of Ukraine, Kharkiv, Ukraine
Mystetstv Street, 4, Kharkiv, 61002, Ukraine

[2]Space Research Centre of the Polish Academy of Sciences, Warsaw, Poland
Bartycka Street, 18A, Warsaw, 00-716, Poland

[*]Corresponding author. e-mail: chechotkin@rian.kharkov.ua;
Contributing authors: odudnyk@cbk.waw.pl; dudnik@rian.kharkov.ua; yakovlev@rian.kharkov.ua

ORCIDs of the author(s):
Dmytro Chechotkin - 0000-0003-1536-3501
Oleksiy Dudnik - 0000-0002-5127-5843
Oleksandr Yakovlev - 0000-0002-4727-7678

**Abstract**

Studying stream interaction regions (SIRs), from their inception and the dynamics of their development, can provide insights into solar–terrestrial connections. Some in-situ instruments on the Solar Orbiter (SolO) space mission are designed to measure solar wind (SW) and interplanetary magnetic field parameters along the flight path. These instruments are ideal for studying the dynamics of SIR evolution at heliocentric distances of 0.28–1.0 AU and with changes in heliolatitude of 0°–33°. To address the challenges of promptly identifying SIRs and predicting their arrival time on Earth, we consider using trigger events from the Radio and Plasma Wave (RPW)/SolO instrument, which are transmitted in telemetry data packages. We suggest that multiple activations of the trigger mode (SBM1 mode) in the RPW instrument over an interval of up to four hours may reflect the fine structure of large-scale events in SW. Such events can serve as markers for the spacecraft's location within the SIR. In this regard, the 2023 analysis revealed that multiple activations of the SBM1 trigger mode throughout the day accounted for more than 50% of the total number of days for which such events were recorded. Of this number, 63% were events when the trigger algorithm was prompted repeatedly within a time interval of up to four hours. A comparison of the registration times of SBM1 trigger events with the SW parameters obtained from the SWA-PAS and MAG instruments showed that repeated activations of the trigger algorithm occurred at the stream interface surface when a high-speed SW stream and a formed compression region were present. We believe that high gradients of changes in SW parameters in this SIR region lead to intense fluctuations in proton density and magnetic field in SW, which onsets the trigger mode. To identify coronal holes that are a potential source for the observed solar wind structure, we propose a set of analytical expressions that enable estimating the position of coronal holes on the solar disk relative to the moment of SIR registration by SolO instruments.

## 1 Introduction

Stream interaction regions (SIRs) are large-scale structures in the solar wind (SW) and influence many processes in the heliosphere. The SIR is understood as a region of compressed SW plasma, which is formed when the fast SW stream, also known as high speed stream (HSS; $V_{SW}$>450 km∕s), emanating from a coronal hole (CH), compresses the slow SW stream ($V_{SW}$<400 km∕s) originating from the streamer belt. Since CHs are long-lived structures, the associated SIR expands in space and corotates with the period of the Sun's revolution around its axis (~ 27 days). In this case, such SW structures are usually called corotating interaction regions (CIRs) (Gosling & Pizzo 1999; Gosling et al. 2001; Jian et al. 2006; Richardson 2018). Due to the temporal variability of SW structures, HSS may disappear without forming CIRs. However, considering that SIRs which are not recurring can still influence the geomagnetic activity, we include them in our analysis. For textual simplicity throughout the manuscript, we use the term SIR to describe both the non-recurring SIRs and the recurring CIRs.

An inherent feature of SIR is the presence of a spatial compressed region. This region includes an accelerated and compressed space region in the slow SW and a decelerated but also compressed region at the leading edge of HSS (Belcher & Davis 1971). When observing SIRs in temporal dynamics, it is common to distinguish stream interface (SI), forward and reverse shock (Jian et al. 2006). Studies of SIR and CIR based on data from the Wind and ACE spacecraft over a 10-year period showed that only 24% of the recorded SIRs had shock (Jian et al. 2006). Of this number, only about 70% had the forward shock and about 24% had the reverse shock. Since the fast and slow SW streams originate from different regions at the Sun with different magnetic field characteristics, the two cannot mix. However, the SIR can be recognized as a rapid growth of density and temperature and increased amplitude of Alfven waves. The application procedure of these parameters for SIRs identification was discussed in more detail by Gosling & Pizzo (1999); Jian et al. (2006); Hajra & Sunny (2022). This procedure assumed monitoring of SW and IMF parameters to search in-situ the presence of characteristic SIR signatures and testing for the presence of high-speed streams from CH. An important factor indicating a correspondence between the CH and HSS is the coincidence of the polarities of the CH and the IMF at the time of in-situ observation of the HSS. Compliance testing of such correspondence can be done using the Helioseismic and Magnetic Imager (the Solar Dynamics Observatory mission) SDO/HMI instrument (Pesnell et al. 2012) and solar synoptic maps from the Space Weather Prediction Center[1] as a subdivision of the National Oceanic and Atmospheric Administration (NOAA SWPC). This technique for detecting SIR can be implemented by considering the near-Earth satellite constellation, as, for example, shown by Hajra & Sunny (2022) when compiling the CIR database during Solar Cycle 24.

SIRs can modulate and scatter galactic cosmic rays (Rouillard & Lockwood 2007) and lead to the acceleration of energetic particles (Heber et al. 1999; Mason & Sanderson 1999; Wijsen et al, 2021; Silwal, et al. 2024). Chi et al. (2018) analyzed the geo-effectiveness of the SIR over the period 1995-2016 and showed that they caused geomagnetic storms of which 52% had a $Dst_{min} \leq -30$ nT and 3% a $Dst_{min} \leq -100$ nT. Zhang et al. (2007) found that during solar cycle 23 only ~10% of the geomagnetic storms with $Dst \leq -100$ nT were associated with SIRs. The SIR-triggered heterogeneity of the Earth's ionospheric plasma in sporadic $E_s$ layers can markedly affect radio communication and navigation systems (Yu et al. 2021). SIRs can often be observed to interact with other SW structures such as slow interplanetary coronal mass ejections (ICMEs), plasmoids formed at the edges of streamer belts, and transient processes associated with the temporal evolution of the boundaries of source regions of fast and slow SW (Jian et al. 2006; Jian et al. 2019; Wen He et al, 2018). When SIR interacts with ICME, disturbances in the form of shock waves, tangential and rotating discontinuities can occur (Jian et al. 2006; Echer et al. 2013; Chi et al. 2018; Heinemann, et al. 2019). This increases the geoeffectiveness of the SIR when it reaches Earth's orbit.

Streamers are one of the sources of slow SW. Unlike pseudostreamers, they separate coronal holes with open field lines of opposite polarity. Dense plasma emanating from streamers forms two layers of heliospheric plasma sheets separated by a heliospheric current sheet (HCS) (Crooker et al. 2014; Tsurutani et al. 2011b). As a result, the HCS appears to be often embedded in the slow SW stream. This leads to a change in the SIR characteristics. Thus, (Jian et al. 2019) found the studied SIRs, that in 54% HCS crossing was observed, and in 34% of cases, the presence of HCS was not observed for 3 days relative to the stream interface on either side. The HCS presence within the SIR structure led to an increase in the proton density and a slight increase in the magnetic field intensity, which caused stronger compression and pressure increase within the stream.

Data on SW and IMF parameters at small distances from the Sun can now be obtained from two solar missions, the Parker Solar Probe (Fox et al. 2016) and Solar Orbiter (SolO) (Müller et al. 2020). The Parker Solar Probe flight path is conducted at heliocentric distances from less than 10 $R_s$ to nearly 1 AU, and SolO flight is conducted at heliocentric distances from 0.28 AU to nearly 1 AU. We use data from the SolO instruments, the Solar Wind Analyser (SWA) (Owen et al .2020), and the Magnetometer (MAG) ([Horbury et al. 2020](#)), and study the possibility of using trigger events from the Radio and Plasma Wave (RPW) instrument ([Maksimovic et al. 2020](#)) as markers for the SolO presence within the SIR. The SolO trigger data are transmitted promptly and are available to the general scientific community within a few days after transmission ,and to special services even earlier, whereas L2 SW and IMF data are only available to the broader scientific community after at least 3 months. Therefore, we investigate whether RPW trigger modes can be used to provide earlier detection of discontinuities and irregularities in the SW, and in particular, SIRs.

---

[1] https://www.swpc.noaa.gov

The Survey Burst Modes 1 (RPW/SBM1 trigger mode) was developed for registering shock waves in the heliosphere. The RPW device has two trigger operating modes, SBM1 and SBM2 (Survey Burst Modes 2) ([Maksimovic et al. 2020](#)). A detailed description of the RPW functioning algorithms in these modes is presented by [Maksimovic et al. (2015)](#). The SBM1 mode enables the registration of interplanetary shock waves (IPS), while the SBM2 mode enables the registration of in-situ type III radio bursts and associated Langmuir waves. However, after a quick analysis of a small number of SBM1 events revealed that the trigger mode is onset not only by shock waves but also by other irregularities (discontinuities) in the SW and magnetic field. Repeated triggers have been found at the leading edge of the HSS, where compression regions are formed. In this work, the SIR is selected for study as one of the large-scale irregularities that can lead to the SBM1 trigger.

The usage of trigger modes allows the monitoring of complex events that require data from various SolO devices. For example, SWA, MAG, and RPW data are used to record shock waves in SBM1 mode, and the Energetic Particle Detector (EPD) suite and RPW data are used to record Langmuir waves in SBM2 mode. The general logic of scientific measurements is that SolO instruments continuously monitor the measured parameters with a high time resolution. The trigger algorithms help select areas corresponding to certain physical phenomena from a continuous stream of data in the region where the trigger is activated, and high-resolution data is recorded in the spacecraft's memory. Information about trigger events is transmitted in a telemetry package. The latter is used to schedule the load of the communication link with scientific data ([Kruparova et al. 2013](#); [Maksimovic et al. 2020](#)).

The basis of the SBM1 trigger mode is an algorithm for detecting IPS of various types. The SBM1 algorithm detects combined temporal jumps in magnetic field strength, proton density, and SW velocity. The algorithm realizes window-time averaging of these parameters over the interval ∆t_buff (currently this parameter is about 3 min). After that, the differences between neighboring time windows are calculated. These differences are then compared with threshold values, above which the decision to trigger is made. [Kruparova et al. (2013)](#) conducted statistical testing of the algorithm's performance for various input data and threshold values. They concluded that the detection of IPS by the algorithm is a probable event. That is why they introduced the so-called quality factor $QF$ which is a generalizing criterion of the functioning of the SBM1 trigger mode algorithm. It is determined by the weighted sum of the averaged fluctuations of the SW and IMF parameters during the analyzed time interval, which is 6 min ([Maksimovic et al. 2015](#); [Maksimovic et al. 2020](#)). By varying the weight factors and the threshold value, one can optimize the operation of the algorithm for specific conditions, for example, heliocentric distance. Thus, for the thresholds applied in the RPW instrument ([Maksimovic et al. 2015](#))—with the threshold for quality factor $QF = 0.2$—the algorithm detected only 60% of IPS from the test list. With the same threshold $QF = 0.2$ and an arbitrary flow of input data, the algorithm detected only 29% of real IPSs, and the remaining detections corresponded to strong fluctuations of plasma and magnetic field parameters or other discontinuities ([Kruparova et al. 2013](#)).

This study aims to assess whether SBM1 events from RPW in January to December 2023 can indicate when SolO was located in SIRs identified from SW parameters obtained from the Proton and Alpha particle Sensor of the SWA (SWA-PAS).

**2 Methodology**

**2.1 Methodology for identifying SIR events**

Jian et al. (2006); Hajra and Sunny (2022); Sunny et al. (2023) use SW and IMF parameters to identify SIRs. In our work, we also use SW parameters (proton density $N_{SW}$; plasma pressure $P_{SW}$; speed $V_{SW}$; temperature $T_{SW}$) and the IMF parameters — IMF magnitude $B_{tot}$ and projections $B_r$, $B_t$, $B_n$ onto the corresponding axes of the RTN coordinate system referenced to SolO (Müller et al. 2020; Marirrodriga et al. 2021).

To identify the SIR, it is necessary to establish the presence of HSS from the CH and the stream interface (Fig. 1(i)). Ideally, the interface is a discontinuity separating the accelerated (pile-up) slow SW and the slowed fast SW followed by an unperturbed stream of high-velocity plasma. Signs of HSS are increased SW velocity to $V_{SW} > 450$ km/s and temperature at low flux densities. The stream interface is characterized by the fact that it balances the dynamic pressure on both sides of the flow and was determined by the maximum of this parameter. Based on this, the boundaries of the SIR were also determined, i.e., a decrease in pressure to the background value (Jian et al. 2006; Jian et al. 2019). Additional criteria for the presence of SIR are increased density, temperature, magnetic field strength, and pressure, which are caused by plasma compression in the region.

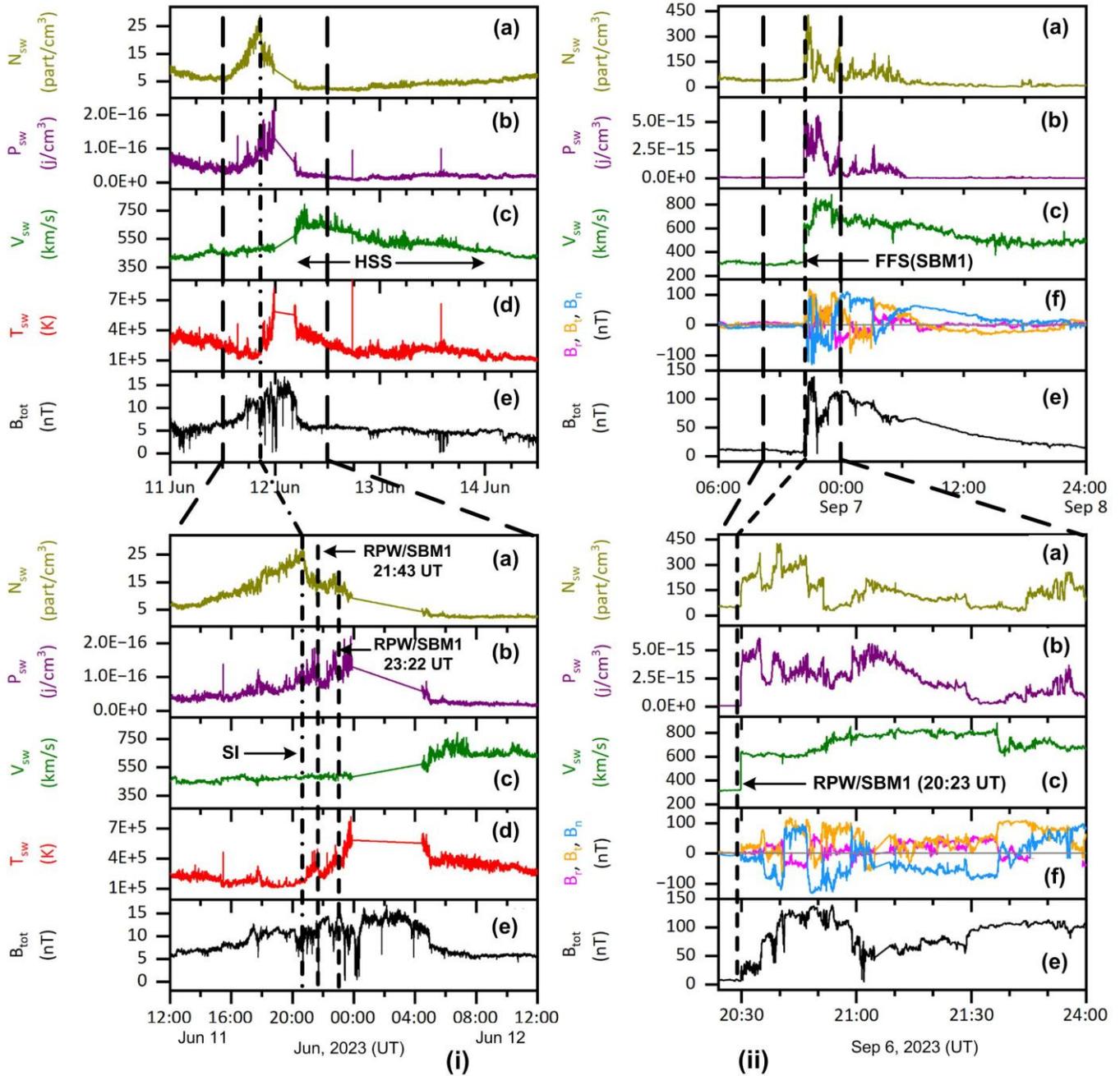

**Figure 1.** Typical behaviour of the SW and IMF parameters during observations with SolO instruments at the moments when the SBM1 is triggered:
(i) HSS and the compression region at its leading edge in the time interval 11.06.2023-14.06.2023, and twice the SBM1 triggers within the compression region (heliocentric distance $D_{SolO-Sun} = 0.89$ AU).

(ii) fast forward shock wave (FFS) at the time of the SBM1 trigger event on 06.09.2023 at 20:23 UT (heliocentric distance $D_{\text{SolO-Sun}} = 0.64$ AU). The driver of this FFS is ICME, since a flux rope is observed in the IMF components after the FFS.

The panels show: (a) proton density $N_{\text{SW}}$; (b) plasma pressure $P_{\text{SW}}$; (c) speed $V_{\text{SW}}$; (d) temperature $T_{\text{SW}}$; (e) IMF magnitude $B_{\text{tot}}$; (f) IMF components $B_r$, $B_t$, $B_n$; RPW/SBM1 are trigger moments.

Next, identifying characteristic SIR elements, such as forward shock, stream interface, and reverse shock, is significant. The presence or absence of specific SIR elements, along with their quantitative characteristics, can vary significantly. This depends on external conditions: other phenomena, such as magnetic clouds and ICMEs, can superimpose on the SIRs and distort their characteristics (Jian et al. 2006; Chi et al 2018). The heliocentric distance from the Sun also affects the SIR characteristics (Gosling and Pizzo 1999; Richardson 2018; Zhong et al. 2024).

The final stage of SIR identification involves searching for CH as a potential HSS source in the solar disk image. To check whether the SIRs mapped back to CHs, we used images of the Sun taken by the SDO/AIA instrument in the range of 21.1 nm, as well as solar synoptic maps[2] NOAA SWPC, and the angular position of the potential HSS source was calculated.

A comparison of the behaviour of the SW and IMF parameters for different physical phenomena at the moments of SBM1 triggering is shown in the Fig. 1. Figure 1(i) shows a case when two SBM1 triggers occurred during passage of an SIR and were not associated with IPSs. A contrasting case of an SBM1 trigger associated with an ICME-driven shock is shown in Fig. 1(ii). We have investigated whether RPW/SBM1 trigger events could be used as potential markers for the presence of SolO within the SIR. For this the SBM1 events were selected based on the following criteria: a) SBM1 events occurred at the leading edge of the HSS; b) SBM1 events could be single events or could have a repeated pattern over several hours; c) SBM1 events did not have clear signs of a fast forward shock.

**2.2 Calculation of the SIR corotation delay between Earth and SolO**

To implement the final part of SIR identification that involves searching for CHs as potential sources of HSS using SDO/AIA images and solar synoptic maps, it is necessary to perform back calculation of the point position on the solar disk relative to the moment of SIR registration at the SolO location. Vennerström et al. (2003) and Opitz et al. (2009) studied different phenomena in SW streams at different points in the heliosphere. They estimated the delay in recording the same event between two points in space by calculating the time lag caused by the difference in the position of these points in space in terms of radial distance from the Sun and the angle in the ecliptic plane at the time of registration of the event at one of the points.

Let us use the assumptions regarding the SIR structure proposed by Chi et al. (2022); Zhong et al. (2024) according to which the formed stable SIR structure rotates perfectly with the Sun, and the leading edge of the SIR can be approximated by the Parker spiral expression. In addition, there are no other structures such as CMEs that interact with the SIR. Taking these assumptions into account, let us write an expression to calculate the time of SIR registration near the Earth relative to the moment of registration by SolO instruments

$$t_{lag} = t_0 + \frac{\Delta\phi}{\Omega - \Omega_E} + \frac{(D_E - D_{\text{SolO}})\Omega}{V_{\text{SW}}(\Omega - \Omega_E)}, \qquad (1)$$

where
$t_0, [s]$ is the initial time at the moment of recording SIR by SolO devices;
$\Delta\phi, [degr]$ is the angle between SolO and the Earth in the ecliptic plane at the time when a SIR encounters SolO,
$V_{\text{SW}}, [km/s]$ is the average radial speed of the SW stream for the observed event;
$\Omega, [degr/s]$ the rotation rate of the Sun in the near-equatorial region,
$\Omega_E, [degr/s]$ is the angular velocity of the Earth's motion relative to the Sun,

---
[2] https://www.predsci.com/mhdweb/spacecraft_mapping.php

$D_{SOLO}$, $[km]$ is the distance between the SolO and the Sun;
$D_E$, $[km]$ is the distance of the Earth to the Sun.

Expression (1) can be simplified with sufficient accuracy for practical application. For this purpose, we take the value $\Omega - \Omega_E = 1.53 \times 10^{-3} [degr/s]$ obtained taking into account the synodic period of the Carrington rotation of the Sun equal to 27.27 days. Since this value of the period corresponds to the direct rotation of the Sun at latitudes $\pm 26^0$ from the equator, which is a characteristic value for the regions of manifestation of periodic solar activity. With the proposed simplification, expression (1) can be simply used to predict the observation time of the SIR structure near the Earth when it is detected by SolO instruments.

**2.3 Estimation of the angular position of the CH on the solar disk at the time of SIR recording**

SDO/AIA imagery and solar synoptic maps were used to confirm CH presence, which could be the sources of each high-speed SW streams that cause the formation of SIR regions. The selection and analysis of SDO/AIA images and solar synoptic maps were adjusted for delay or advance relative to the analyzed event at a specific angle to the central meridian.

To estimate the angular position of the CH on the solar disk relative to the central meridian at the time of SIR registration by SolO instruments, we use the Parker spiral expression from Zhong et al. (2024) for a polar coordinate system (R, ϕ). Assuming that CH, as a potential source of HSS, is located in the equatorial region (this precision is sufficient to detect CH on EUV images of the solar disk), and the origin of the coordinate system coincides with the Sun-Earth direction. The geometry of the problem on the ecliptic plane is shown in Fig. 2.

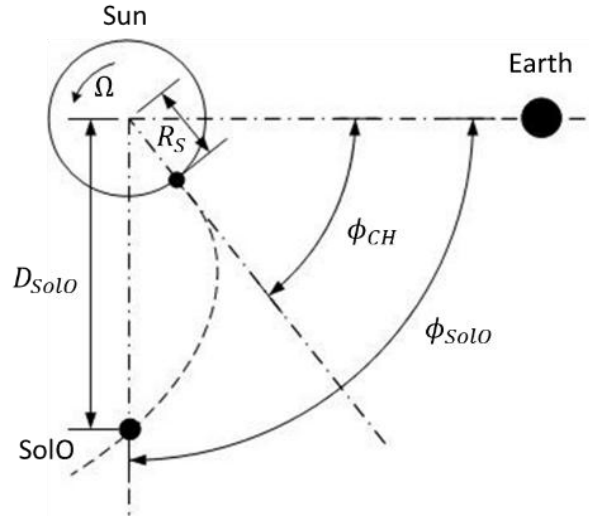

**Figure 2.** Geometry of calculating the position of the CH relative to the central meridian at the moment of SIR registration by SolO instruments.

Then, the Parker spiral equation for the case shown in Fig. 2 can be written as

$$D_{SolO} - R_S = \frac{-V_{SW}}{\Omega}(\phi_{SolO} - \phi_{CH}), \qquad (2)$$

where
$D_{SolO}$ is the distance to SolO from the coordinate origin,
$R_S$ is the Sun's radius,
$\Omega$ is the angular velocity of the Sun's rotation in the equatorial region,
$V_{SW}$ is the average radial speed of the SW stream for the observed event,
$\phi_{SolO}$ is the angular position of SolO relative to the Earth.
$\phi_{CH}$ is the angular position of CH relative to the central meridian for the location of Earth.

From (2), the position of the CH relative to the central meridian at the moment of event registration by SolO instruments can be estimated by the meridian angular shift:

$$\phi_{CH} = \phi_{SolO} - \frac{\Omega}{v_{sw}}(R_S - D_{SolO}). \quad (3)$$

Applying expression (3) has the following requirements: i) if the angular position of SolO is eastward relative to the central meridian, then the value of $\phi_{SolO}$ in expression (3) should be substituted with the minus sign; ii) if the positive result is obtained, the potential CH is located to the west of the central meridian and to the east if the negative result is obtained.

To search for the required SDO/AIA image in the database, it is more convenient to operate not with the angular advance (delay) in degrees, but with a similar parameter in days. For this purpose it is more convenient to use the expression

$$\phi_{CH}[day] = -0.075\, \phi_{CH}[degree],$$

where $\phi_{CH}[degree]$ is the value of the angular advance (delay) in degrees obtained on the basis of expression (3). Then, if the result of calculations is positive, we search for the required image in the SDO/AIA database with an advance by $\phi_{CH}[day]$ relative to the event registration date, and if the result is negative - with a delay by $\phi_{CH}[day]$ relative to the event registration date.

Some of the events we considered were complex events of interaction between ICME and SIR. In such situations, we used the DONKI database and in-situ data from the SWA and the MAG instruments to search for a CME that potentially encountered SolO.

## 3 Results

### 3.1 Statistical analysis of SBM1 events for 2023

Considering the peculiarities of the operation of the SBM1 algorithm in the RPW instrument and the probabilistic nature of shock event registration using the algorithm described by Kruparova et al. (2013), we systematized all SBM1-type events registered throughout 2023. Table 1 presents statistics of SBM1 events for each month of 2023.

Trigger thresholds of the SBM1 algorithm were optimized to identify forward, reverse, fast, and slow shock waves (Maksimovic et al. 2015). FFS waves are often observed at the ICME front, propagating at speeds exceeding 500 km/s (Tsurutani et al. 2011a). Considering the behavior of the SW and IMF parameters at the leading edge of the SIR (if a forward shock has formed) (Gosling & Pizzo 1999; Richardson 2018), we can conclude that the forward shock of the SIR is FFS. Therefore, the FFSs were analyzed separately and are presented in Table 1. FFS was identified visually by detecting an intensive increase in SW and IMF parameters, included in the SBM1 algorithm (Kruparova et al. 2013) — the parameters considered are the IMF intensity, SW speed, and SW proton density (Fig. 1ii). The decision about the presence of FFS was made if the parameter jump occurred within a time window of 3 min, which is used in the SBM1 algorithm (Maksimovic et al. 2015).

**Table 1.** Statistics of SBM1 events throughout 2023. $N_{SBM1}$ is the total number of SBM1 events per month; $N_{FFS}$ is the number of SBM1 events that are FFS per month; $N_{day}$ is the number of days in the current month when at least one SBM1 event was registered; $N_{day-rep}$ is the number of days in a month in which two or more SBM1 events were recorded; $N_{day-4hh}$ is the number of days in a month in which repeating SBM1 events were recorded with an interval of no more than 4 hours; $D_{Sun-SolO}$ is the distance between Sun-SolO at the beginning, middle, and end of each month, respectively.

| Month | $N_{SBM1}$ | $N_{FFS}$ | $N_{day}$ | $N_{day-rep}$ | $N_{day-4hh}$ | $D_{Sun-SolO}$, AU | | |
|---|---|---|---|---|---|---|---|---|
| | | | | | | 1st day of | 15th day of | 30th day of |

| | | | | | | | the month | the month | the month |
|---|---|---|---|---|---|---|---|---|---|
| 1 | 18 | 1 | 13 | 4 | 4 | | 0.95 | 0.95 | 0.92 |
| 2 | 13 | 2 | 8 | 3 | 1 | | 0.91 | 0.84 | 0.73 |
| 3 | 39 | 2 | 9 | 4 | 4 | | 0.72 | 0.57 | 0.37 |
| 4 | 124 | - | 25 | 19 | 7 | | 0.35 | 0.31 | 0.50 |
| 5 | 42 | 1 | 13 | 10 | 8 | | 0.51 | 0.67 | 0.82 |
| 6 | 12 | 1 | 9 | 3 | 2 | | 0.83 | 0.90 | 0.94 |
| 7 | 24 | 2 | 14 | 5 | 3 | | 0.95 | 0.95 | 0.92 |
| 8 | 23 | 2 | 13 | 6 | 4 | | 0.91 | 0.83 | 0.70 |
| 9 | 7 | 3 | 5 | 2 | 1 | | 0.69 | 0.51 | 0.35 |
| 10 | 6 | 1 | 4 | 1 | 1 | | 0.33 | 0.35 | 0.54 |
| 11 | 14 | 5 | 9 | 3 | 2 | | 0.56 | 0.72 | 0.83 |
| 12 | 29 | 3 | 11 | 8 | 6 | | 0.84 | 0.91 | 0.95 |
| ∑ | 351 | 23 | 133 | 68 | 43 | | | | |
| % | | 5.7% of the SBM1 events during the year | | 51.1% of the sum of $N_{day}$ | 63.2% of the days with repeated SBM1 events ($N_{day\text{-}rep}$); | | | | |

The total number of FFS ($N_{FFS}$), which may correspond to events such as forward shock SIR, ICME, or their interaction, was about 6% of the total number of burst mode triggers during 2023. SolO completed two full orbits around the Sun during this time. Some studies show an increase in the number of IPS occurs at a distance greater than 0.6 AU (e.g., Dimmock et al. 2023). This may be due to the fact that fast forward shocks generated at the leading edges of SIRs are less likely to be found closer to the Sun. In this study, we focused on FFSs and did not count the number of other types IPSs (slow and reverse shock). We assume that the result obtained does not contradict the results of laboratory testing of the algorithm—the number of IPSs detected was 29% at $QF = 0.2$ (Kruparova et al. 2013). In three months (4, 5 and 12) SBM1 events occurred two or more times on at least 50% of days in the month when SBM1 events were recorded. On more than 63% of these days, several SBM1 events were recorded within a time interval of less than 4 hours.

**3.2 High-speed SW streams and SBM1 events in January 2023**

Using the example of one of the months presented in Table 1, we will consider in more detail the timestamps an SBM1 was registered at relative to the SW parameters. During January 2023, the trajectory of SolO relative to the Earth and the Sun by longitude and distance was within small limits and amounted to $D_{SolO-Sun} = 0.92 - 0.95$ AU and $L_{SolO-S-E} = 22° - 30°$ east of the Earth.

In Fig. 3, vertical dotted lines correspond to SBM1 events. The total number of vertical lines corresponds to the $N_{day}$ parameter from Table 1. At each vertical line, the FFS event and clustered events $N_{day\text{-}4hh}$, indicating the number of SBM1 triggers on that day, were marked. Despite missing data for some days, SBM1 events are observed against the background of four HSS lasting from 3 to 5 days. One $N_{FFS}$ event was recorded on 17.01.2023 at 09:56 UT. This appears to correspond with an IPS that was detected by STEREO-A/IMPACT instruments at 14:32 UT according to the DONKI database (STEREO A was close to SolO, ~13 deg. west of SolO and 0.02 AU further from the Sun). The source of this event may be a shock wave at the CME front. This IPS may be associated with one of the CMEs that occurred in AR13191 on 15.01.2023 at 03:48 UT and 05:00 UT, respectively, and registered in the DONKI database. These CMEs were observed on the visible side of the solar disk in the southeast and were confirmed by LASCO-C2 images. Following this event there is a gap in the SWA/PAS data, after which, according to the SW speed data, a continuation of the HSS is observed (Fig. 3c).

Some SBM1 events are distinctively grouped in areas of peak density, pressure, IMF intensity, and maximum speed for HSS, for example, on the following days: 01.01.2023; 02.01.2023; 04.01.2023; 08.01.2023; 09.01.2023; 13.01.2023; 17.01.2023; 24.01.2023; and 30.01.2023. Noteworthy is the fact that, for some days (09.01.2023; 13.01.2023; 30.01.2023), repeated activations of the SBM1 algorithm were observed in the areas of peak density and pressure of the SW stream.

As noted by Gosling & Pizzo (1999) and Jian et al. (2006), the pressure gradient controls the evolution of SW structures and can be informative for SIR identifying. At the same time, the pressure parameter is not used in the SBM1 algorithm (Kruparova et al. 2013; Maksimovic et al. 2007). Therefore, we monitored pressure dynamics in conjunction with other SW parameters and illustrated this in our figures.

From the example shown in Fig. 3, we can conclude that repeating SBM1 events for up to 4 hours tend to occur against a background of large-scale structures in the SW. If these events occur at local peaks of pressure and density of SW when HSS is present, they can indicate that SolO is traversing a SIR.

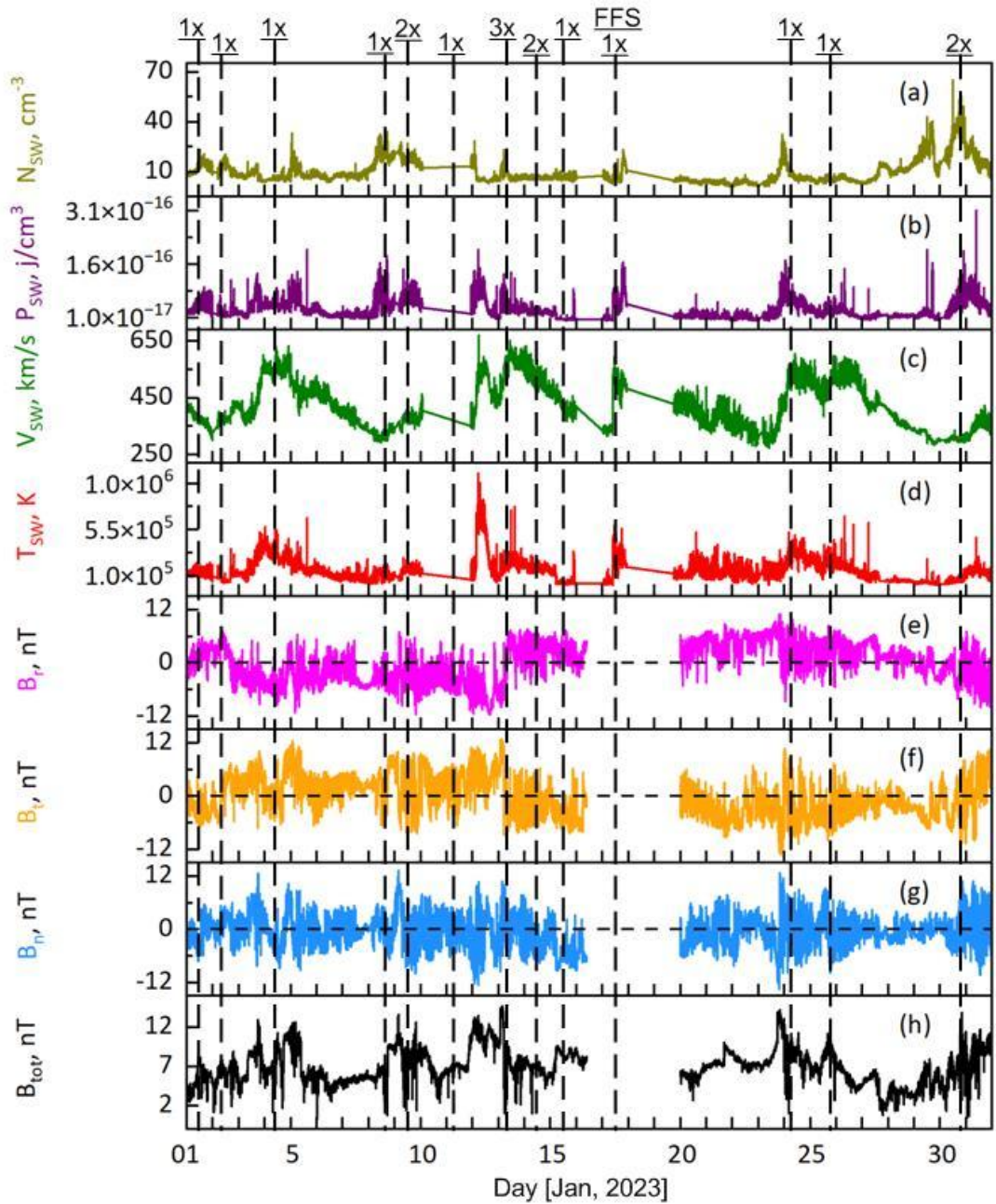

**Figure 3.** Times of SBM1 events versus the background of the dynamics of SW and IMF parameters throughout January 2023: (a) proton density $N_{SW}$; (b): pressure $P_{SW}$; (c): speed $V_{SW}$; (d): temperature $T_{SW}$;

(e): IMF components $B_r$, $B_t$, $B_n$; (f): IMF magnitude $B_{tot}$; vertical dashed lines indicate times of registrations of SBM1events corresponding to $N_{day}$ (Table 1); $N$x is the number ($N$) of repeated SBM1 events on this day.

## 3.3 Selected examples of SBM1 events occurring during different SW and IMF discontinuities

### 3.3.1 Event 1: series of SBM1 triggers on January 13, 2023

Let us consider in more detail the HSS interval from January 13, 2023, to January 15, 2023, shown in Fig. 3, and the adjacent regions. At this time, the distance of SolO from the Sun was $D_{SolO-Sun} = 0.95$ AU, the angle in the ecliptic plane was $L_{SolO-S-E} = 26°$ east of the central meridian. Following January 15, 2023, there is a gap in the data (approximately 23 hours on January 16, 2023), but based on the observed trend in the SW velocity (Fig. 3c), it can be assumed that HSS was observed for about 5 days, approximately until the beginning of January 17, 2023.

At the leading edge of the HSS, between 00:30-10:00 UT January 13, 2023, a compression region can be observed, identified by a jump in density and pressure in the SW stream (Fig. 3a, b). This region corresponds to a triple SBM1 trigger. Further to the left, there is a disturbed section in the SW. Based on the intense fluctuation of the SW and IMF parameters (Fig. 3) throughout 12.01.2023, it can be assumed that this transient event is not associated with HSS.

To the left of the disturbed region, there is no SWA-PAS data, and the MAG data shows a region of weak and slow SW. The change in the IMF sectors occurred in the compression region on January 13, 2023. This can be seen from the behavior of the IMF components (Fig. 3e, f, g). The slow SW section is observed in the negative polarity sector, and the adjacent HSS is in the positive polarity sector of the IMF. Consequently, the HCS crossing occurs in the compression region.

Figure 4 shows in detail the compression region, the leading edge of the HSS, and the triggering moments of the three SBM1 triggers on January 13, 2023. We define the boundaries of the compression region on the left by FFS at 01:40 UT, and on the right by the return of the SW pressure to the background value, following the recommendations of Jian et al. (2019).

Despite the FFS presence at the leading edge of the SIR, no SBM1 events were recorded. We believe that the growth of SW parameters in this section exceeded the trigger threshold values; however, the rate of their increase was too low for the trigger to be reliably onset.

Three SBM1 events were recorded in the center of the compression region, clearly visible on the SWA-PAS spectrogram (Fig. 4(i)). Figure 4(i) shows the variation of the density (coloured scale on the right) and velocity, which are plotted as energy equivalent (on the Y vertical axis), simultaneously for protons and alpha-particles with time. This spectrogram demonstrates the presence of the HSS and the compression region at its leading edge as indicated by the variations in particle densities..

The first two SBM1 triggers at 05:00 UT and 06:14 UT occur against the backdrop of similar behavior in SW and IMF parameters (Fig. 4) – a local maximum in SW density and pressure against a backdrop of a local depression in temperature and magnetic field intensity on an hourly scale. Such behavior of parameters is characteristic at the moments of HCS crossing (Liou & Wu 2021). At SBM1 trigger moments, a change in the polarity of the radial and tangential components of the IMF is observed (Fig. 4e, 4f).

Liou & Wu (2021) and Khabarova et al. (2021a) note the possibility of multiple HCS crossings within the SIR. Khabarova et al. (2021a) found that between the start of the SIR and the SI, there is an almost twofold increase in the number of HCS crossings, with a peak at the SI location, followed by a gradual decrease. They suggest that this peak in the SI region is associated with the level of turbulence observed downstream, where SI represents the strongest discontinuity at 1 AU. In our case, an additional contribution was made by a transient event in the slow SW stream, which could have played a role in the shift of HCS towards SI.

The third trigger, SBM1 at 07:41 UT, also coincides with local maxima in density and pressure in the SW stream. Since this is followed by a steady decrease in pressure and density to background values, while velocity and temperature reach maximum values, we associate this SBM1 event with SI.

Taking into account the mutual position of SolO, the Earth, and the Sun and using expression (3), we obtain that the angular position of HSS relative to the central meridian is $\phi_{CH} \approx 15^0$. Thus, at the moment of SIR registration by SolO instruments, the HSS source on the solar disk may be located in the equatorial area and slightly west of the central meridian. Figure 5a shows the SDO/AIA image in the 21.1 nm band on the day of the event registration by the SolO instruments. The HSS source was surrounded by active regions 13182, 13184, 13186, and 13187 and in this perspective slightly complicates the identification of the CH (indicated by the arrow in Fig. 5a). At the same time, on the solar synoptic map[3], we can observe this CH in the area of positive polarity, which corresponds to the direction of the radial component of the IMF (Fig. 4e) at the time when the HSS reached its maximum speed.

---

[3] https://www.ngdc.noaa.gov/stp/space-weather/solar-data/solar-imagery/composites/full-sun-drawings/boulder/2023/01/boul_neutl_fd_20230113_0646.jpg

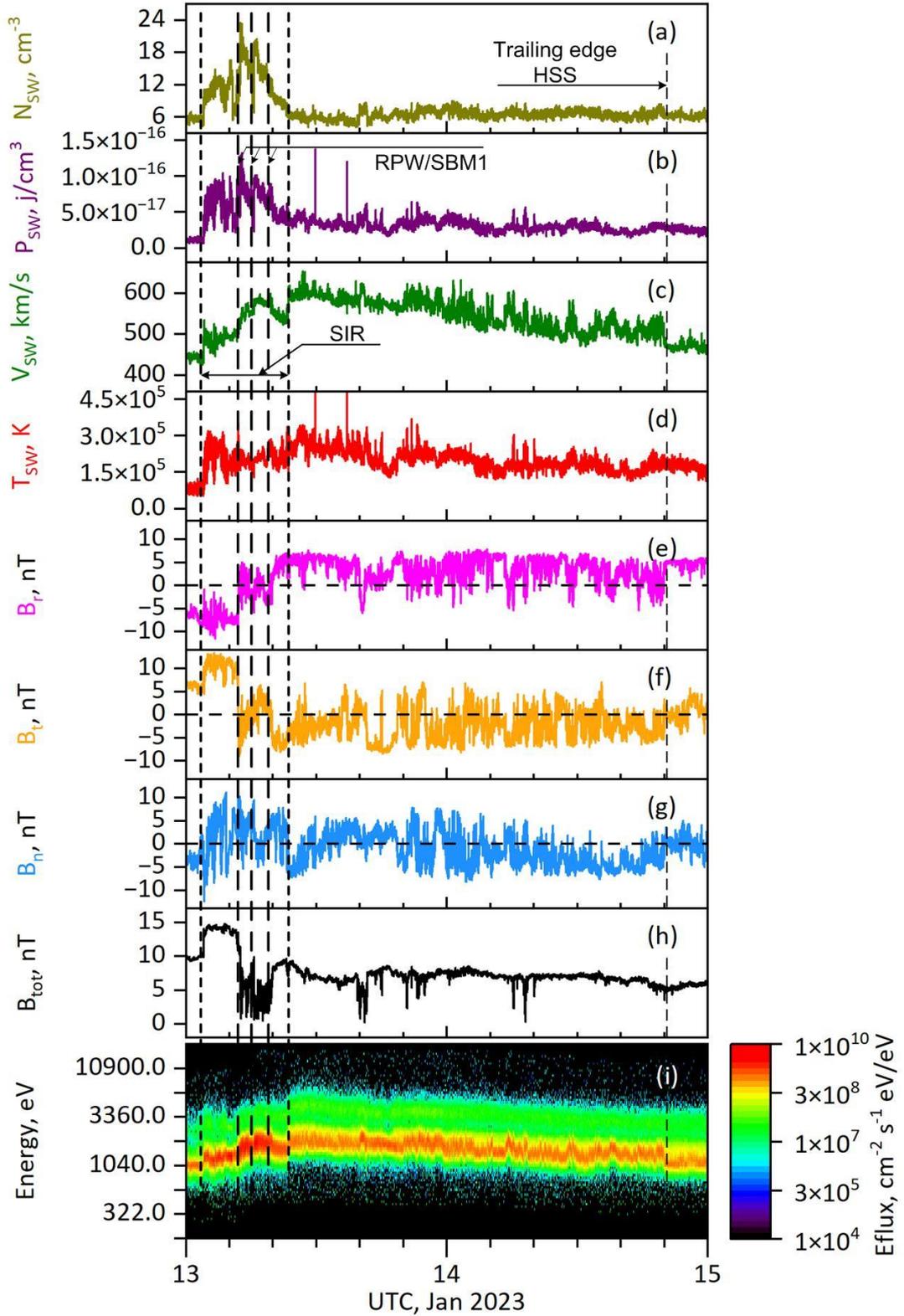

**Figure 4.** The changes in SW and IMF parameters on January 13–14, 2023: (a) proton density $N_{SW}$; (b) pressure $P_{SW}$; (c) speed $V_{SW}$; (d) temperature $T_{SW}$; (e) IMF components $B_r$, $B_t$, $B_n$; (f) IMF magnitude $B_{tot}$; (g) dynamic spectrograms of SW protons and alpha-particles.

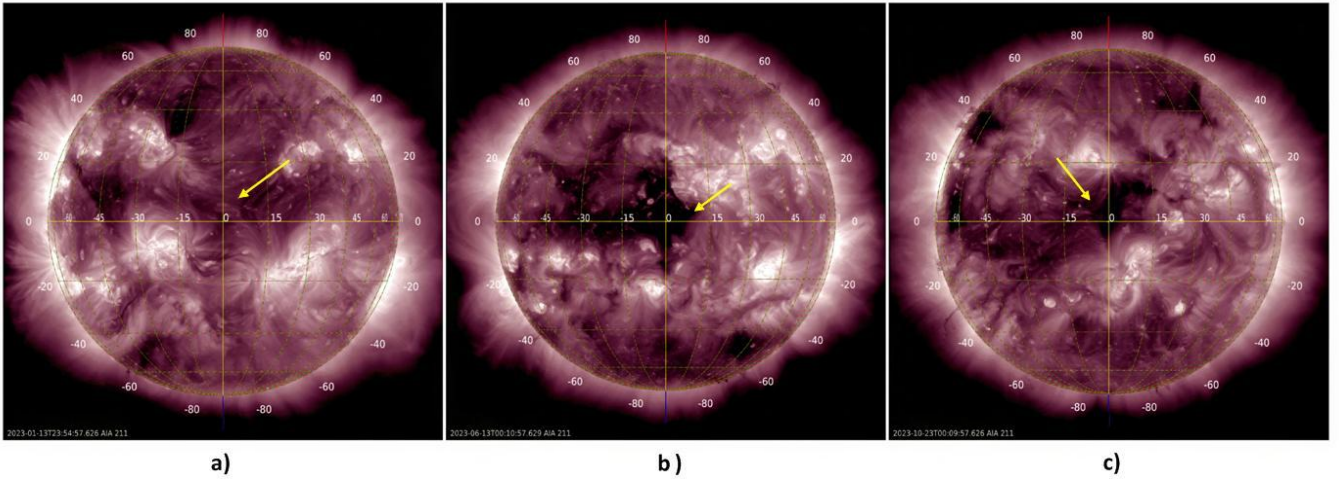

**Figure 5.** SDO/AIA images at 21.1 nm demonstrate the position of CH on the solar disk as a potential HSS source for the events under consideration (see the text for more details): a) image from 13.01.2023 / 23:55 UT for event 1, b) image from 13.06.2023 / 00:11 UT for event 2, c) image from 23.10.2023 / 00:10 UT for event 3. The arrow points to the location of the potential HSS source.

### 3.3.2 Event 2: series of SBM1 triggers on June 28, 2023

Let us consider a series of two SBM1 triggers that were detected at the leading edge of the HSS on June 28, 2023 (Fig. 6). Fig. 6j shows the entire HSS and the behavior of the SW and IMF parameters that preceded it and may influence the interpretation of events. Figure 6(jj) shows the leading edge of the HSS in more details, including the compression region and the moments of SBM1 triggering. The formed compression region was visible on the SWA-PAS spectrogram (Fig. 6(i)) in the region of the SBM1 events with a further smooth transition to a rarefied region and high-speed stream. After 12:00 UT on June 29, 2023, a rarefaction area was observed in the HSS. This event was special because SolO was on the far side of the Sun with respect to Earth. The SolO position relative to the Earth and the Sun at that moment was $D_{\text{SolO-Sun}} = 0.94$ AU, $L_{\text{SolO-S-E}} = 161°$ west of the central meridian.

As can be seen in Fig. 6(j), the HSS had a short duration, slightly more than 1.5 days, and reached a maximum speed of about 650 km/s. Judging by the density and pressure parameters (Fig. 6a, 6b), a compression region formed at the leading edge of the HSS. To the left of it is a disturbed region, which we interpret as an ICME (shown in Fig. 6(j)), as fragments of the ICME can be observed in the IMF components. The beginning of the disturbed region occurs at 00:15 UT on June 26, 2023, at the moment of the FFS-type SBM1 trigger. Based on the behavior of the curves in Fig. 6(j), we cannot confirm the presence of FFS, because after the trigger, there is an approximately six-hour section of data gap.

Due to the presence of a disturbed region, it is difficult to clearly define the left boundary of the compression region. Following the recommendations Jian et al. (2006) and Jian et al. (2019), we define the boundaries of this region based on the behavior of pressure and density (Fig. 6a, b), from the beginning of growth and return to background values. To the left of the disturbed region, there is a calm region of slow wind. It is noteworthy that the calm region of slow SW and HSS has different IMF polarities. The boundary between magnetic fields of opposite polarity is referred as the sector boundary (Crooker et al. 2004). Sectors of opposite polarity are separated by HCS, which is the largest and most stable structure of the SW in the heliosphere. HCSs are generated at the cusp of the streamer belt and are an element of slow SW (Liou & Wu 2021). Transient events can lead to local deformation of this structure and complicate the signatures for identification.

At around 10:00 UT on June 28, 2023, RPW detected two consecutive SBM1 events with an interval of about one hour (Fig. 6(jj)). Both SBM1 triggers occurred against the background of similar behavior of the SW and IMF parameters—a local maximum in SW density and pressure accompanied by a local depression in IMF intensity over a period of several hours. A few hours after the second SBM1 trigger, the radial and tangential components of the IMF (Fig. 4e, 4f) stabilized. This indicates a change in IMF sectors, and throughout the HSS, we observe the sector with polarity pointing toward the Sun (in contrast to the section of calm, slow SW in Fig. 6(j), where we observe the sector with polarity pointing away from the Sun). Given the above, we suggest that the change in IMF sectors, and consequently the intersection of HCS, occurred in the compression region, and SBM1 triggers may indicate the moment of this crossing.

Taking into account the relative positions of SolO, Earth, and Sun, and using expression (3), we obtain that the angular position of the HSS source relative to the central meridian at the time of SBM1 event registration will be $\phi_{CH} \approx 202^0$. To select a suitable SDO/AIA image from the database, we convert the calculated value in degrees of the angular position of the HSS source into a delay in days $\phi_{CH} \approx -15$[day], caused by the rotation of the calculated point on the solar surface to the central meridian. Figure 5b shows the SDO/AIA image from June 13, 2023, where we can observe the near-equatorial CH near the central meridian. Additionally, the solar synoptic map[4] shows that this CH has negative polarity, which coincides with the polarity of the radial component of the IMF at the time of the HSS observation in Fig. 6e

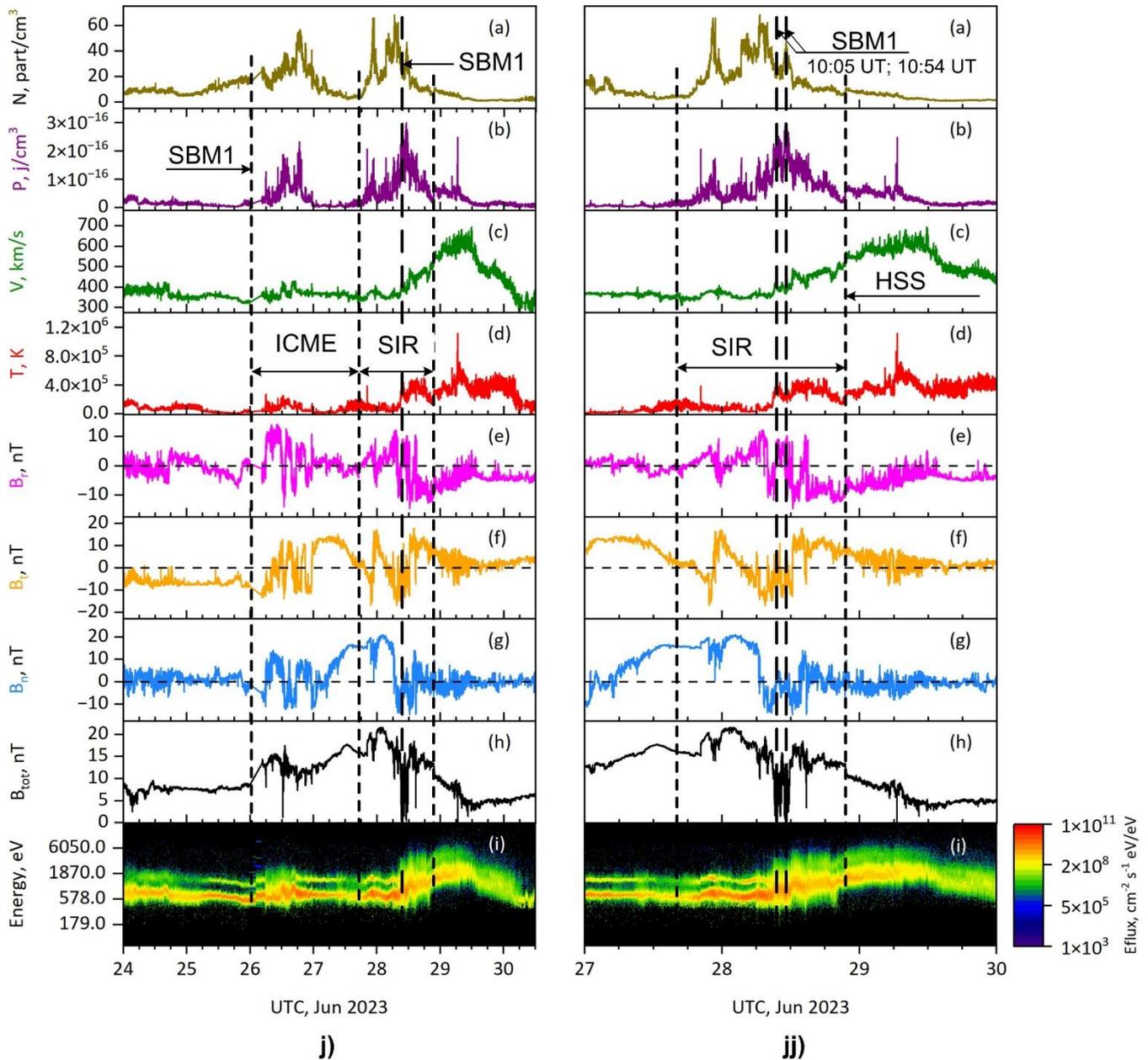

**Figure 6.** The changes in SW and IMF parameters at the time of registration of a series of SBM1 triggers on June 28, 2023: (a): proton density $N_{SW}$; (b): pressure $P_{SW}$; (c): speed $V_{SW}$; (d): temperature $T_{SW}$; (e): IMF components $B_r$, $B_t$, $B_n$; (f): IMF magnitude $B_{tot}$; (g): dynamic spectrograms of SW protons and alpha-particles.

---

[4] https://www.ngdc.noaa.gov/stp/space-weather/solar-data/solar-imagery/composites/full-sun-drawings/boulder/2023/06/boul_neutl_fd_20230613_1316.jpg

### 3.3.3 Event 3: series of SBM1 triggers on October 24, 2023

There's a common feature between the two events we considered and discussed in Sections 3.3.1 and 3.3.2. HSS and the preceding compression region of the slow SW stream were located in opposite sector boundaries of the IMF. And despite the presence of a disturbed region immediately before the SIR, the change in magnetic field sectors occurs in the compression region and coincides with the SBM1 trigger moments.

Next, we will examine a series of three SBM1 triggers that were detected at the leading edge of the HSS on October 24, 2023 (Fig. 7). Figure 7(j) shows the HSS duration of about 4 days (October 25-28), the compression region at its leading edge (October 23-24), and the preceding HSS to the left. Figure 7(jj) shows the leading edge of the HSS in more detail, including the compression region and the moments when the SBM1 triggers were activated. The presence of the compression region is detected by areas of increased density and pressure (Fig. 7a, 7b).

Although the IMF components are variable (Fig. 7e, 7f, 7g), it can be concluded that the HSS under study and the HSS that preceded it, as well as most of the compression region, are located in the sector with the same IMF anti-sunward polarity. At the same time, the sector boundary between the magnetic field sectors and the HCS has not yet formed. During the phase of solar activity growth, the dipole structure of the solar magnetic field is destroyed, and multipole components begin to dominate. During the years of the solar maximum, the dipole heliomagnetic field is replaced by quadrupole and octupole fields, and even higher harmonics (Maiewski et al. 2020). Therefore, during periods close to the maximum of solar activity, compression regions that do not include IMF sector boundaries occur more frequently. As shown by the studies of Crooker et al. 2014, regardless of the source of the slow SW stream, other characteristics of the compression region are very similar.

At the beginning of October 24, 2023, RPW detected three SBM1 events (Fig. 7(jj)). The first two SBM1 triggers, at 00:41 UT and 01:30 UT, demonstrate similar behavior to the SW and IMF parameters namely a local maximum in SW density and pressure accompanied by a local depression in temperature and IMF intensity with a change in the polarity of the radial and tangential components (Fig. 7(jj)). This behavior of the parameters lasts for several hours, after which they return to their previous values. At 05:33 UT, the third SBM1 trigger is detected, with signatures similar to the previous two and a duration of IMF depression of about half an hour.

When describing the trigger moments of the previous two events, we noted that the triggering occurred at the moment of HCS crossing, but in the period from October 21 to October 29, 2023, no IMF sector changes were observed. Khabarova et al. (2021b) and Pezzi et al. (2021) note that Current Sheets (CSs) are formed everywhere in the heliosphere and have different spatial scales. From thin CSs, several proton gyroradius thick, to large and quasi-stable CSs, to which the authors also include HCS. Thin, or local, CSs are formed at the boundaries of plasmas with dramatically different properties, discontinuities, and separatrices. Studying the frequency of CSs occurrence in ACE data for the period 1998-2012 (Khabarova et al. 2021a) found that in the most turbulent regions of the SIR, the number of CSs can reach hundreds per hour. We suggest that all three SBM1 triggers occurred at local CSs in the SIR, as turbulence areas in the SW stream are increased.

The position of SolO relative to the Earth and the Sun on October 23, 2023, was $D_{\text{SolO}-\text{Sun}} = 0.45$ AU, $L_{\text{SolO}-\text{S}-\text{E}} = 32°$ east of the central meridian. Considering the relative position of SolO and using Expression (3), we obtain that the angular position of the CH relative to the central meridian is $\phi_{\text{CH}} \approx -5°$ and, at the time of SIR registration, the HSS source was located almost at the center of the solar disk (Fig. 5c).

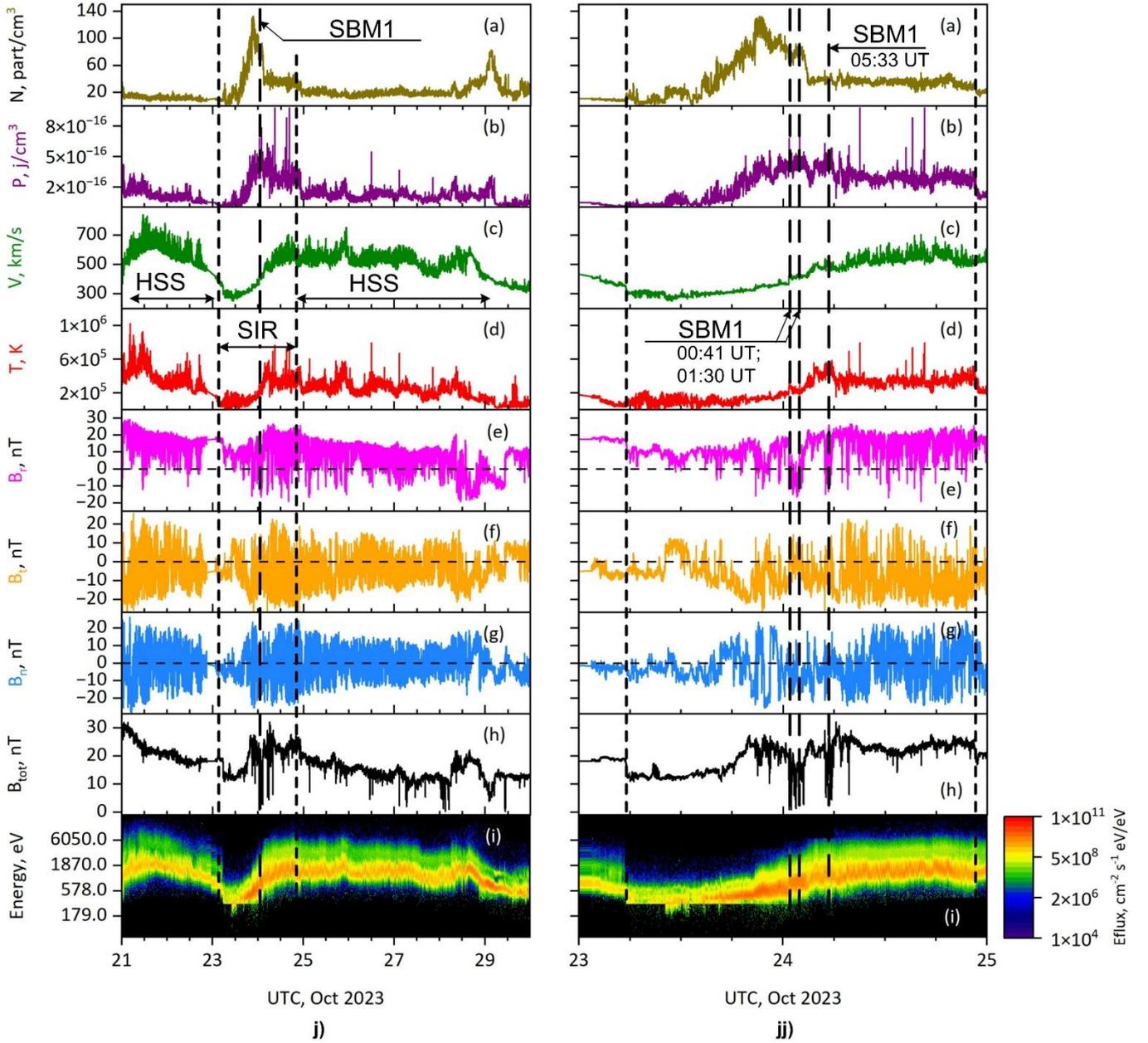

**Figure 7.** The changes in SW and IMF parameters at the time of registration of a series of SBM1 triggers on October 24, 2023: (a): proton density $N_{SW}$; (b): pressure $P_{SW}$; (c): speed $V_{SW}$; (d): temperature $T_{SW}$; (e): IMF components $B_r$, $B_t$, $B_n$; (f): IMF magnitude $B_{tot}$; (g): dynamic spectrograms of SW protons and alpha-particles.

## 4 Discussion and conclusions

Analysis of SBM1 events for the year 2023 showed that there were fewer FFS-related triggers (about 6%) than laboratory studies of the algorithm (about 29%) conducted by Kruparova et al. (2013) had shown. Maksimovic et al. (2021) note that not all SBM1 events indicate IPS, but they do point to discontinuities in the SW stream that require further investigation. These authors proposed a method for using SBM1 triggers that do not show IPS characteristics to identify parallel whistlers in the vicinity of the discontinuity (Maksimovic et al. 2021). As a follow-up, we suggest using SBM1 triggers that do not have FFS signatures and repeat at short time intervals to detect compression regions.

We looked at what happened with SBM1 in 2023 and found that in 50% of cases, the trigger is triggered several times per day. In 63% of such days, the re-triggering occurs with an interval of up to 4 hours. Taking into account that large-scale structures in the heliosphere have a temporal extent of more than a day, such triggers may characterize internal changes in these structures and correspond to their separate elements, for example, forward and reverse shocks, stream interfaces, and others.

As the events reviewed have shown, the SBM1 trigger is activated multiple times by fluctuations in the SW and IMF parameters caused by the crossing of the CS, which is often located in the compression region. Moreover, the trigger is activated both when the HCS crosses with a change in IMF sectors and when the local CS crosses, caused by increased turbulence of the SW stream in the compression region. We assume that the trigger activation in these sections is caused by a short-term and intense depression of the magnetic field intensity and a local peak in proton density, which is a typical signature of the CS.

For the events examined, the time interval between several consecutive registrations of SBM1 triggers when crossing CS in the compression region ranged from 50 minutes to 2 hours 40 minutes. This is compatible with the hypothesis of multilayered current structures in HCS and is associated with the fact that streamers can have a rather complex internal structure with several neutral surface layers ([Maiewski et al. 2020](#)).

The observational data suggest that repeated SBM1 events, which lack clear signs of an FFS wave and are recorded over two to four hours, can indicate the SolO presence within the SIR. More reliable conclusions regarding the registration and characteristics of SIR can be drawn after studying the parameters of SW and the interplanetary magnetic field from data obtained from the SWA and MAG. Data on SW and IMF parameters from these instruments are made available to the scientific community with a three-month delay, and a brief report on SBM1 trigger events is transmitted in telemetry packets. This allows SBM1 events to be used for prompt analysis of discontinuities and irregularities in the SW stream. And, as the results show, for identifying SIRs and predicting their arrival time on Earth.

**Acknowledgements**

This work is supported by the "Long-term program of support of the Ukrainian research teams at the Polish Academy of Sciences carried out in collaboration with the U.S. National Academy of Sciences with the financial support of external partners", Agreement No. PAN.BFB.S.BWZ.363.022.2023.

The authors express their gratitude to the scientific and technical group of the SDO/AIA Mission, RPW, SWA, and MAG teams of the Solar Orbiter Mission, the NOAA Space Weather Prediction Center and the Community Coordinated Modeling Center and the Moon to Mars Space Weather Analysis Office at the NASA Goddard Space Flight Center for free access to the data.

The authors acknowledge Prof. C.J. Owen from MSSL-UCL at University College London, UK, and prof. T. Horbury from the Blackett Laboratory at Imperial College London, UK for providing the data from SWA and MAG.

The authors would like to thank Enago (www.enago.com) for the English language review.

**Data availability**

The work used data available in open sources.
1. Solar Orbiter/SWA, Solar Orbiter/RPW, Solar Orbiter/MAG data are available at: https://soar.esac.esa.int/soar/#home
2. SDO/AIA data is available at: https://sdo.gsfc.nasa.gov/data/aiahmi/
3. Current space weather conditions on NOAA Scales. Solar Synoptic Map: https://www.swpc.noaa.gov/products/solar-synoptic-map
4. SOHO/LASCO data is available at: https://soho.nascom.nasa.gov/data/archive.html
5. The Space Weather Database of Notifications, Knowledge, Information (DONKI): https://ccmc.gsfc.nasa.gov/tools/DONKI/